\documentclass[twocolumn,showpacs,epj,referee]{svjour}

\usepackage{graphics}% Include figure files
\usepackage{latexsym}

\newcommand{\abs}[1]{\left\vert#1\right\vert}
\newcommand{\ket}[1]{\left\vert#1\right\rangle}
\newcommand{\bra}[1]{\left\langle#1\right\vert}

\begin{document}

\title{Spatial decoherence factor via the qubit-field interaction}
\titlerunning{Spatial decoherence factor}
\author{A. Vaglica  and G. Vetri}
\authorrunning{A. Vaglica and G. Vetri}
\institute{%
CNISM and Dipartimento di Scienze Fisiche ed Astronomiche,
Universit\`{a} degli Studi di Palermo, via Archirafi
36, 90123 Palermo, Italy.}%

\date{\today}

\abstract{We analyze the time evolution of an initial spatial
coherence for a two level atom whose internal degrees of freedom
interact with a single mode of a cavity field. When the
qubit-field subsystem is taken as an environment, the
translational dynamics experiences a decoherence process which may
be encoded in a decoherence factor $D$. We find that the field
statistics affects $D$ through the alternative paths the
system-environment may follow along their entanglement, while
eventual field phase properties give rise to an imaginary part of
$D$ which is related to the atomic translation. From the
decoherence perspective, we analyze the relation between the
atomic momentum and the imaginary part of the atomic spatial
density matrix, and some considerations on its asymptotic behavior
are brought into question at the conclusion of the paper.}
\PACS{{03.65.Yz} Decoherence; open systems; quantum statistical
methods \and {37.10.Vz} Mechanical effects of light on atoms,
molecules, and ions}
\maketitle
\section{\label{sec:level1} Introduction}

Decoherence program \cite{ze,zu1,zu2,schlo1,joos,schlo2} may be
considered a serious attempt to overcome the basic dilemma of
Quantum Mechanics (QM): Why the world appears as classical,
despite of its underlying quantum nature that allows for arbitrary
superpositions of states? An implication inherent to this program
is the possibility of dividing the world into subsystems. As
suggested by Zeh \cite{land,schlo1}, any observation involves
ignorance of a subsystem (a part of the universe) and this
procedure defines the ''facts'' that can be realized in a quantum
system. Accordingly, Landsman \cite{land,schlo1} asserts that a
measurement, a fact or event in QM implies the non-observation, or
irrelevance, of a certain part of the system in question.

Remarkable experiments on the decoherence of mesoscopic coherent
fields in a cat state and complementarity experiments on Rydberg
atoms interacting with microwave cavities have been performed by
Brune et al. \cite{bru} (see also the review of Raimond, Brune,
and Haroche \cite{haro}). Similar foundational aspects of
complementarity, which-way information and quantum erasure have
been studied by Storey, Collett, Walls \cite{SCW1,SCW2}, Scully,
Englert, Walther \cite{SEW}, Storey, Tan, Collett, Walls
\cite{STCW}, D\"{u}rr, Nonn, Rempe \cite{DNR}, in systems implying
interactions and correlations among atomic internal and external
dynamics and laser or maser cavities.

It is usually understood that the irrelevant subsystem works as a
reservoir, that is, it consists of the innumerable degrees of
freedom of the environment with which the system of interest
interacts. However, in a natural way one may extend the analysis
to environments with a few degrees of freedom (see for example
\cite{hu,av,vv}) and to pay attention to the role that the density
of the states the irrelevant subsystem can effectively accede to,
plays in the decoherence process. The study of these cases in
which the system may follow a relatively small number of
alternatives, can be useful for a grasp in the quantum-classical
transition. For instance, non local correlations of entangled
Bell's states may be suddenly and quite irreversibly destroyed by
an environmental action of a subsystem of continuous one-degree of
freedom \cite{vv}. Coherences and non local correlations are in
fact very sensitive to any environmental action, and the ensuing
information leakage towards the irrelevant part of the system is
at the origin of the classical landing.\\
Decoherence effects on qubits caused by the atomic motion and,
viceversa, effects of the qubit-field interaction on the atomic
spatial coherences have been considered in different contests (see
for example \cite{av,vv,you,shre,tum,zhe,tum1,tum2,chia}). In
particular, in Ref. \cite{tum2,chia} the optical Stern-Gerlach
(SG) model was used to analyze complementarity, which-path
information and quantum erasure. Using the same model, the time
behavior of the entanglement between the internal dynamics of a
two-level atom, the zero point cavity field and the transverse
translational variables of the atomic center of mass, was
analytically studied in Ref. \cite{tum}. Disregarding the
translational dynamics by tracing on the relative variables, the
field-qubit coherences go to zero at long time. The diagonal form
asymptotically attained by the reduced density matrix, clearly
indicates how the field-qubit system goes towards the
separability. In addition, if one choose the qubit as the system
of interest, that is, tracing also on the field variables, one
trivially recover the classical diagonal form of the qubit density
matrix \cite{vvv}. These are two examples of decoherence in which
a decisive role is played by the irrelevance of a single degree of
freedom, with a continuum of accessible states, that is, the
conjugate variables that describe the atomic kinematics along the
cavity axis.

In the present paper we will inquire on the atomic spatial
coherences caused by the qubit-field interaction. Using the same
model, we analyze the time evolution of
an initial spacial coherence under the effect of the interaction
between the field of an ideal cavity and the atomic internal
dynamics. Looking at the qubit-field variables as the irrelevant
part of our system, a decoherence effect of the atomic position
follows, and we show that the density matrix that describes the
translational dynamics at a generic time $t$ may be simply
factorized in terms of the initial density matrix and a
decoherence factor, $\rho(x,x';t)=\rho(x,x';0)D(x,x';t)$, where
$D(x,x';t)$ will depend on the field statistics. We analyze the
decoherence process for different initial configurations of the
qubit-field subsystem. Starting from the more general state (pure
or mixed) of the cavity single mode, we then specialize to some
cases, including the thermal and coherent states. We also give the
results for the so-called eigenstate of the Sussking-Glogower
phase operator \cite{SussCarr,Cir,cus}. In all these cases we find
a loss of coherences of the atomic position, similar to the well
known phenomenon of the Rabi oscillations collapse in the usual
Jaynes-Cummings model \cite{naro,rempe,kni,riti,bar}. Because of
the discreteness of the variables that account for our
environment, a partial revival of the coherences follows at
relatively long times.

Finally, we will address to the question of the decoherence
process in the presence of a mean value of the atomic momentum. We
find that for an increasing mean atomic momentum, as it is for
some configurations here considered,  the imaginary part of
$\rho(x,x';t)$ does survive in some regions of the plane $(x,x')$.
\section{\label{sec:level2} Interaction of a travelling qubit with
a single cavity mode}
To pick up the spatial decoherence effect solely caused by the
entanglement with the qubit-field subsystem, we will consider
Rydberg atoms interacting with microwave cavity field. The same
kind of interaction is used by Scully et al. \cite{SEW} in their
analysis of complementarity, based on matter-wave interferometry.
The possibility of realizing initial coherent spatial
distributions using Rydberg atoms in microwave cavities, has
actually been called in question in Ref. \cite{SCW2}. On the other
hand, strong-coupling conditions readily obtained for these atoms
and high quality factors $Q$, allow to neglect, with some
accuracy, both spontaneous atomic emission ($\tau_{a}\sim 10^{-2}
sec$) and cavity loss ($\tau_{c}\sim 10^{-3} sec$) \cite{haro} for
the atomic flight times we will use to obtain full decoherence
($T_{0}\leq 10^{-3} sec$).
\subsection{\label{sec:level2} Model and initial configuration}
The optical SG model is particularly suitable for an analytical
study of our subject. It consists, as known, of the usual
Jaynes-Cummings model in which, however, the dynamics of the
atomic center of mass along the cavity axis is taken into account,
\begin{equation}\label{ham}
\hat{H}=\frac{\hat{p}^{2}}{2\,m}+\hbar \omega
\left(\hat{a}^{\dag}\hat{a}+\hat{S}_{z}+\frac{1}{2}\right)+ \hbar
\varepsilon k
\hat{x}\,(\hat{a}^{\dag}\hat{S}_{-}+\hat{a}\,\hat{S}_{+}),
\end{equation}
and correlates with the dynamics of the other subsystems. This
Hamiltonian describes, in the rotating wave approximation, the
resonant interaction of a two-level atom of mass $ {m} $ with the
resonant $ {k} $-mode of an ideal cavity. It is supposed that the
atomic flight time inside the cavity is sufficiently small to
treat classically the degree of freedom of the atomic center of
mass along the direction orthogonal to the cavity axis. On the
contrary, the atomic transverse dynamics in the ${x}$-direction,
along the cavity axis, is quantized and initially described by a
packet of width narrow with respect to the wavelength $ {\lambda}
$ of the resonant mode, and centered near a nodal region of the
sinusoidal mode function. In these conditions it may be
approximated by the linear term, as the interaction part of Eq.
(\ref{ham}) shows. The conjugate variables $\hat{x} $  and $
\hat{p} $ just describe this transverse dynamics, $ \hat{a} $ and
$ \hat{a}^{\dag}$ are the usual field operators, and $ \varepsilon
$ is the atom-field coupling constant. Finally, the $ 1/2 $ spin
operators $ \hat{S}_{z}$ and
$\hat{S}_{\pm}=\hat{S}_{x}\pm\imath\hat{S}_{y}$ account for the
qubit dynamics.

The evolution operator for the Hamiltonian (\ref{ham}) may be
factorized \cite{cus}, for example, in the  following form,
\begin{eqnarray}\label{evol1}
  \hat{U}(t,0)=\exp\left\{-\frac{2it}{\hbar}\,m\,\hat{a}_{N}\,\hat{\mu}_{x}\hat{x}\right\}
  \exp\left\{-\frac{it}{2m\hbar}\hat{p}^{2}\right\}\nonumber \\
  \times\exp\left\{\frac{i}{\hbar}\,\hat{a}_{N}\,{t}^{2}\hat{\mu}_{x}\hat{p}\right\}
  \,e^{-i\,\vartheta_{0}(t)\,\hat{N}},\qquad  t\leq{T}_{0}
\end{eqnarray}
where $ {T}_{0} $  indicates the atomic flight time inside the
cavity, and
\begin{eqnarray}
\hat{a}_{N}={a}_{0}\,\sqrt{\hat{N}},\,\,\,{a}_{0}=\frac{\varepsilon\,\hbar\,k}{m},\,\,\,
\vartheta_{0}(t)=\omega\,t+m\,{a}_{0}^{2}\,t^{3}/6\,\hbar \label{acc1}\\
\hat{N}=(\hat{a}^{\dag}\hat{a}+\hat{S}_{z}+\frac{1}{2}),\qquad\qquad
\hat{\mu}_{x}=\frac{\hat{a}^{\dag}\hat{S}_{-}+\hat{a}\,\hat{S}_{+}}{2\sqrt{\hat{N}}}.\,\label{acc2}
\end{eqnarray}
We suppose that the initial state of the entire system is given by
\begin{equation}\label{rotqf}
\hat{\rho}(0)=\hat{\rho}_{t}(0)\hat{\rho}_{q}(0)\hat{\rho}_{f}(0)
\end{equation}
where
\begin{equation}\label{rotq}
\hat{\rho}_{t}(0)=\ket{\varphi(0)}\bra{\varphi(0)},\qquad\hat{\rho}_{q}(0)=\ket{\varphi_{q}(0)}\bra{\varphi_{q}(0)}
\end{equation}
account for the initial configuration of the atomic external and
internal degrees of freedom, respectively, while
\begin{equation}\label{rof}
\hat{\rho}_{f}(0)=\sum_{n,n'}c_{n,n'}\ket{n}\bra{n'}
\end{equation}
describes a generic (pure or mixed) state of the cavity field,
expressed in terms of Fock states $\ket{n}$. The qubit state
$\ket{\varphi_{q}(0)}$ is a coherent superposition
\begin{equation}\label{phiq}
\ket{\varphi_{q}(0)}=\cos\frac{\gamma}{2}\ket{e}+e^{i\phi}\sin\frac{\gamma}{2}\ket{g},\qquad
0\leq\gamma\leq\pi
\end{equation}
of excited $\ket{e}$ and ground $\ket{g}$ states.

We are interested to the behavior of the atomic spatial
coherences, so we consider a coherent superposition of two nearly
distinct kets \cite{zu1,zhe},
\begin{equation}\label{superpos}
\ket{\varphi(0)}=
\frac{1}{\sqrt{2\delta}}[\ket{\varphi_{1}(0)}+\ket{\varphi_{2}(0)}]
\end{equation}
to describe the initial position distribution of the atomic center
of mass along the cavity axis. In particular, we will assume that
the ${x}$-representation of $\ket{\varphi_{j}(0)}$ ($j=1,2$) is
given by Gaussian functions,
\begin{equation}\label{gauss}
\varphi_{j}(x,0)=\left(\frac{1}{\sqrt{2\pi}\Delta
x_{0}}\right)^{\frac{1}{2}}
\exp\left\{-\frac{(x-x_{0,j})^2}{4\Delta x_{0}^{2}}\right\},
\end{equation}
centered in $x_{0,1} $ and $x_{0,2} $, respectively, around the
origin of the reference frame which is set in a nodal point of the
mode function. The normalization constant of state
(\ref{superpos}),
\begin{equation}\label{delta}
\delta=\left[1+\exp\left\{-\frac{(x_{0,1}-x_{0,2})^{2}}{8\Delta
x_{0}^{2}}\right\}\right],
\end{equation}
accounts for the eventual (small) overlap of the two Gaussians.
For simplicity, the Gaussian distributions (\ref{gauss}) are of
minimum uncertainty, $\Delta x_{0}\Delta p_{0}=\hbar/2$, with the
same widths $\Delta x_{0}$ and $\Delta
p_{0}$ for both the Gaussians.\\
\subsection{\label{sec:level2} Time evolution of the full density operator}
Let us consider the density operator
$\hat{\chi}(0)\equiv\hat{\rho}_{q}(0)\otimes\hat{\rho}_{f}(0)$ of
the irrelevant subsystem. In terms of the dressed states
\begin{equation}\label{vest}
\ket{\chi_{n}^{\pm}}=\frac{1}{\sqrt{2}}[\ket{e,n}\pm\ket{g,n+1}],\,\,\,\ket{g,0},
\end{equation}
it assumes the following form
\begin{eqnarray}\label{rochi0}
\hat{\chi}(0)=\frac{1}{2}\sum_{n,n'=0}^{\infty}\left\{A_{n,n'}\ket{\chi_{n}^{+}}\bra{\chi_{n'}^{+}}\right.
\qquad\qquad\qquad\qquad\nonumber\\
\left.+B_{n,n'}\ket{\chi_{n}^{-}}\bra{\chi_{n'}^{-}}\right.
\left.+C_{n,n'}\ket{\chi_{n}^{+}}\bra{\chi_{n'}^{-}}\right\}\nonumber\\
+\frac{1}{2}\sum_{n=0}^{\infty}
\left\{D_{n}\left(\ket{\chi_{n}^{+}}+\ket{\chi_{n}^{-}}\right)\bra{g,0}\right.\qquad\qquad\qquad\nonumber\\
\left.+E_{n}\left(\ket{\chi_{n}^{+}}-\ket{\chi_{n}^{-}}\right)\bra{g,0}\right\}
+\frac{F}{2}\ket{g,0}\bra{g,0}\nonumber\\
+h.c.\qquad\qquad\qquad\qquad\qquad\qquad\qquad\qquad\qquad
\end{eqnarray}
where
\begin{eqnarray}
A_{n,n'}=\frac{1}{2}\left\{c_{n,n'}\cos^{2}(\gamma/2)+\left[c_{n,n'+1}e^{-i\phi}\right.\right.\qquad\qquad\qquad\nonumber\\
\left.\left.+c_{n+1,n'}e^{i\phi}\right]
\cos(\gamma/2)\sin(\gamma/2)
+c_{n+1,n'+1}\sin^{2}(\gamma/2)\right\}\,\,\,\,\,\,\label{coeffAnn}\\
B_{n,n'}=\frac{1}{2}\left\{c_{n,n'}\cos^{2}(\gamma/2)-\left[c_{n,n'+1}e^{-i\phi}\right.\right.\qquad\qquad\qquad\nonumber\\
\left.\left.+c_{n+1,n'}e^{i\phi}\right]
\cos(\gamma/2)\sin(\gamma/2)
+c_{n+1,n'+1}\sin^{2}(\gamma/2)\right\}\,\,\,\,\,\,\label{coeffBnn}\\
C_{n,n'}=\left\{c_{n,n'}\cos^{2}(\gamma/2)-\left[c_{n,n'+1}e^{-i\phi}-c_{n+1,n'}e^{i\phi}\right]\right.
\,\,\,\,\,\nonumber\\
\left.\times\cos(\gamma/2)\sin(\gamma/2)
-c_{n+1,n'+1}\sin^{2}(\gamma/2)\right\}\label{coeffCnn}\qquad\\
D_{n}=\sqrt{2}\cos(\gamma/2)\sin(\gamma/2)c_{n,0}e^{-i\phi},\qquad\qquad\qquad\qquad\label{coeffDn}\\
E_{n}=\sqrt{2}\sin^{2}(\gamma/2)c_{n+1,0},\qquad\qquad\qquad\qquad\qquad\qquad\label{coeffEn}\\
F=\sin^{2}(\gamma/2)c_{0,0}.\qquad\qquad\qquad\qquad\qquad\qquad\qquad\,\,\,\,\,\,\,\label{coeffF}
\end{eqnarray}
We note that the dressed states (\ref{vest}) are eigenstates of
the observables $\hat{N}$ and $\hat{\mu}_{x}$ which appear in the
expression (\ref{evol1}),
\begin{equation}\label{autost}
\hat{\mu}_{x}\ket{\chi_{n}^{\pm}}=\pm\frac{1}{2}\ket{\chi_{n}^{\pm}},\,\,\,\,
\hat{N}\ket{\chi_{n}^{\pm}}=(n+1)\ket{\chi_{n}^{\pm}}. \\
\end{equation}
Applying the evolution operator (\ref{evol1}) to the initial state
$\hat{\rho}(0)=\hat{\rho}_{t}(0)\otimes\hat{\chi}(0)$ and using
Eq.s (\ref{superpos}), (\ref{rochi0}) and (\ref{autost}), we get
the state of the entire system at time $t\leq{T}_{0}$,
\begin{eqnarray}\label{rot}
\hat{\rho}(t)=\frac{1}{4\delta}\sum_{n,n'=0}^{\infty}e^{-i\vartheta_{0}(t)(n-n')}\left\{A_{n,n'}
\left[\ket{\phi_{n,1}^{+}(t)}\right.\right.\nonumber\\
\left.\left.+\ket{\phi_{n,2}^{+}(t)}\right]\left[\bra{\phi_{n',1}^{+}(t)}+\bra{\phi_{n',2}^{+}(t)}\right]
\otimes\ket{\chi_{n}^{+}}\bra{\chi_{n'}^{+}}\right.\nonumber\\
\left.+B_{n,n'}
\left[\ket{\phi_{n,1}^{-}(t)}+\ket{\phi_{n,2}^{-}(t)}\right]\right.\qquad\qquad\qquad\nonumber\\
\left.\times\left[\bra{\phi_{n',1}^{-}(t)}+\bra{\phi_{n',2}^{-}(t)}\right]
\otimes\ket{\chi_{n}^{-}}\bra{\chi_{n'}^{-}}\right.\qquad\nonumber\\
\left.+C_{n,n'}
\left[\ket{\phi_{n,1}^{+}(t)}+\ket{\phi_{n,2}^{+}(t)}\right]\right.\qquad\qquad\qquad\qquad\nonumber\\
\left.\times\left[\bra{\phi_{n',1}^{-}(t)}+\bra{\phi_{n',2}^{-}(t)}\right]
\otimes\ket{\chi_{n}^{+}}\bra{\chi_{n'}^{-}}\}\right.\qquad\nonumber\\
+\frac{1}{4\delta}\sum_{n=0}^{\infty}e^{-i\vartheta_{0}(t)(n+1)}\left\{D_{n}\left[
\left(\ket{\phi_{n,1}^{+}(t)}+\ket{\phi_{n,2}^{+}(t)}\right)\ket{\chi_{n}^{+}}\right.\right.\nonumber\\
\left.\left.+\left(\ket{\phi_{n,1}^{-}(t)}+\ket{\phi_{n,2}^{-}(t)}\right)\ket{\chi_{n}^{-}}\right]
\bra{g,0}\left(\bra{\varphi_{1}(t)}+\bra{\varphi_{2}(t)}\right)\right.\nonumber\\
\left.+E_{n}\left[\left(\ket{\phi_{n,1}^{+}(t)}+\ket{\phi_{n,2}^{+}(t)}\right)\ket{\chi_{n}^{+}}
-\left(\ket{\phi_{n,1}^{-}(t)}\right.\right.\right.\nonumber\\
\left.\left.\left.+\ket{\phi_{n,2}^{-}(t)}\right)\ket{\chi_{n}^{-}}\right]
\bra{g,0}\left(\bra{\varphi_{1}(t)}+\bra{\varphi_{2}(t)}\right)\right\}\qquad\nonumber\\
+\frac{1}{4\delta}
F\ket{g,0}\bra{g,0}\otimes\left[\ket{\varphi_{1}(t)}+\ket{\varphi_{2}(t)}\right]
\left[\bra{\varphi_{1}(t)}+\bra{\varphi_{2}(t)}\right]\nonumber\\
+ h.c.\qquad\qquad\qquad\qquad\qquad\qquad\qquad\qquad\qquad
\end{eqnarray}
where
\begin{equation}\label{phit}
\ket{\varphi_{j}(t)}=\exp\left\{-i\frac{\hat{p}^{2}t}{2m\hbar}\right\}\ket{\varphi_{j}(0)}\,\,\,\,\,\,\,
\end{equation}
describes the free evolution of the atomic center of mass for zero
excitations in the qubit-field state, (last term of Eq.
(\ref{rochi0})), while
\begin{eqnarray}\label{phitnpm}
\ket{\phi_{n,j}^{\pm}(t)}=\exp\left\{\mp\frac{i}{\hbar}ma_{n}t\hat{x}\right\}
\exp\left\{-\frac{it}{2m\hbar}\hat{p^{2}}\right\}\nonumber\\
\times\exp\left\{\pm\frac{i}{2\hbar}a_{n}t^{2}\hat{p}\right\}
\ket{\varphi_{j}(0)}\qquad\qquad
\end{eqnarray}
are the scattered components of the spatial atomic states due to
the optical SG effect. In fact, $a_{n}=a_{0}\sqrt{n+1}$ accounts
for the acceleration along the cavity axis of these components. In
the $x-$representation one has
\begin{eqnarray}\label{phinjpmxt}
\phi_{n,j}^{\pm}(x,t)=\left(\frac{\Delta x_{0}}{\sqrt{2
\pi}\beta(t)}\right)^{\frac{1}{2}}\exp\left(\mp \frac{i}{\hbar}
m\,a_{n}\,x\,t\right)\nonumber\\
\times\exp\left\{-\frac{[x-x_{n,j}^{\pm}(t)]^{2}}{4\,\beta(t)}\right\}\qquad\qquad
\end{eqnarray}
with
\begin{equation}\label{xntbetat}
x_{n,j}^{\pm}(t)=x_{0,j}\mp{a_{n}}t^{2}/2,\,\,\,\,\,\,\,\,\,\,
\beta(t)=\Delta x_{0}^{2}+i\hbar t/2m.
\end{equation}
For each value of $n$ we get an uniformly accelerated Gaussian
distribution, with the typical width of a free particle. This
behavior of the width is a consequence of the linearity of the
positional potential energy which appears in the interaction term
of the Hamiltonian (\ref{ham}).
\section{\label{sec:level2}  Decoherence Factor}
As said, in our case the irrelevant part is the qubit-field
subsystem. In other words, we suppose to measure the atomic
position whatever the values the qubit-field variables take on.
From a mathematical point of view, this implies a tracing of the
density operator $\rho(t)$ on the field-qubit variables (for a
discussion on the relation between a reduced density operator and
the actual measurement of an observable, see
\cite{pessoa,schlo1}). In this section we give the exact
expression of the reduced density matrix that describes the atomic
translational dynamics. In addition, we will see that under
sufficiently wide conditions, it is possible to single out a
decoherence factor.
\subsection{\label{sec:level2} Spatial density operator}
Using the orthonormality of the dressed states (\ref{vest}) we get
the following reduced density operator describing the atomic
translation degree of freedom along the cavity axis,
\begin{eqnarray}\label{rored}
\hat{\rho}_{space}(t)=Tr_{field,qubit}[\hat{\rho(t)}]\qquad\qquad\qquad\qquad\nonumber\\
=\frac{1}{2\delta}\sum_{n=0}^{\infty}\left\{A_{n,n}\left[\ket{\phi_{n,1}^{+}(t)}+\ket{\phi_{n,2}^{+}(t)}
\right]\right.\nonumber\\
\left.\times\left[\bra{\phi_{n,1}^{+}(t)}+\bra{\phi_{n,2}^{+}(t)}\right]
+B_{n,n}\right.\qquad\nonumber\\
\left.\times\left[\ket{\phi_{n,1}^{-}(t)}+\ket{\phi_{n,2}^{-}(t)}\right]\left[\bra{\phi_{n,1}^{-}(t)}
+\bra{\phi_{n,2}^{-}(t)}\right]\right\}\nonumber\\
+\frac{1}{2\delta}
F\left[\ket{\varphi_{1}(t)}+\ket{\varphi_{2}(t)}\right]
\left[\bra{\varphi_{1}(t)}+\bra{\varphi_{2}(t)}\right],
\end{eqnarray}
where we have used the fact that the coefficients $A_{n,n},
B_{n,n}$ and $F$ are real. To analyze the atomic spatial
coherences, we take the matrix elements of $\hat{\rho}_{space}(t)$
with respect to the position eigenstates. Using Eq.s (\ref{phit})
and (\ref{phinjpmxt}), and separating the real and the imaginary
parts of the exponents we get
\begin{eqnarray}\label{roredxx'}
\hat{\rho}(x,x';t)\equiv\bra{x}\hat{\rho}_{space}(t)\ket{x'}\qquad\qquad\qquad\qquad\qquad\nonumber\\
=\frac{F}{2\delta\sqrt{2\pi}\Delta
x_l(t)}\sum_{j,k=1}^{2}\exp\left[i\alpha_{0}^{j,k}(x,x')\right]\qquad\qquad\nonumber\\
\times\exp\left\{-\frac{1}{4\Delta x_{l}^{2}(t)}
\left[\left(x-x_{0,j}\right)^{2}+\left(x'-x_{0,k}\right)^{2}\right]\right\}\nonumber\\
+\frac{1}{2\delta\sqrt{2\pi}\Delta x_l(t)}\sum_{n=0}^\infty
\sum_{j,k=1}^{2}A_{n,n}\exp\left[i\alpha_{n,+}^{j,k}(x,x')\right]\nonumber\\
\times\exp\left\{-\frac{1}{4\Delta x_{l}^{2}(t)}
\left[\left(x-x_{n,j}^{+}(t)\right)^{2}+\left(x'-x_{n,k}^{+}(t)\right)^{2}\right]\right\}\nonumber\\
+\frac{1}{2\delta\sqrt{2\pi}\Delta x_l(t)}\sum_{n=0}^\infty
\sum_{j,k=1}^{2}B_{n,n}\exp\left[i\alpha_{n,-}^{j,k}(x,x')\right]\nonumber\\
\times\exp\left\{-\frac{1}{4\Delta x_{l}^{2}(t)}
\left[\left(x-x_{n,j}^{-}(t)\right)^{2}+\left(x'-x_{n,k}^{-}(t)\right)^{2}\right]\right\}\nonumber\\
\end{eqnarray}
%%%\end{widetext}
%
where we have set
\begin{eqnarray}
\alpha_{0}^{j,k}(x,x')=\frac{\hbar t}{8m\Delta x_{0}^{2}\Delta
x_{l}^{2}(t)}\qquad\qquad\qquad\qquad\nonumber\\
\times\left[\left(x-x_{0,j}\right)^{2}-\left(x'-x_{0,k}
\right)^{2}\right]\label{alfa0}\qquad\\
\alpha_{n,\pm}^{j,k}(x,x')= \frac{\hbar t}{8m\Delta
x_{0}^{2}\Delta
x_{l}^{2}(t)}\left\{\left[x-x_{n,j}^{\pm}(t)\right]^{2}\right.\nonumber\\
\left.-\left[x'-x_{n,k}^{\pm}(t)\right]^{2}\right\}\mp\frac{1}{\hbar}ma_{n}t(x-x')\label{alfan}
\end{eqnarray}
and we have used
\begin{eqnarray}
\Delta x_{0}^{2}\Delta
p_{0}^{2}=\hbar^2/4\qquad\qquad\qquad\qquad\qquad\qquad\qquad\label{delta}\\
\abs{\beta(t)}^{2}=\Delta x_{0}^{2}(\Delta x_{0}^{2}+\Delta
p_{0}^{2}t^2/m^2)\equiv\Delta x_{0}^{2}\Delta
x_{l}^{2}(t).\label{beta}\,\,\,
\end{eqnarray}
Once again, we wish to stress that the Eq. (\ref{roredxx'}) is the
exact reduced density operator describing the atomic dynamics
along the cavity axis in the optical SG model, for generic states
of the cavity field and the qubit. In the next section the
decoherence process hidden in Eq. (\ref{roredxx'}) will be
analyzed for some specific states of the cavity field.
\subsection{\label{sec:level2} Decoherence factor}
We now observe that the linear approximation of the cavity mode
function requires small displacement and spread of the atomic wave
packet, small with respect to the wavelength $\lambda$. In other
words, when the atom leaves the cavity its spatial distribution
must be essentially unchanged with respect to the initial one. As
a consequence, the flight time $T_{0}$ must be sufficiently short
to allow the disregard of both the displacement $a_{n}T_{0}^2/2$
and the distortion $\Delta p_{0}T_{0}/m$ in Eq.s (\ref{xntbetat})
and (\ref{beta}). For the values of the parameters used in this
paper, these conditions are fulfilled for $T_{0}\Omega\leq
10^{3}$, where $\Omega=\varepsilon\sqrt{<n>+1}$ is the Rabi
frequency. For flight time satisfying this condition we have (see
Eq.s (\ref{xntbetat}) and (\ref{beta})) $x_{n,j}^{\pm}(t)\simeq
x_{0,j}$ and $\Delta x_{l}^{2}(t)\simeq \Delta x_{0}^{2}$.
Consequently, Eq.s (\ref{alfa0}) and (\ref{alfan}) become
\begin{eqnarray}\label{alfaa}
\alpha_{0}^{j,k}(x,x')\simeq \frac{\hbar t}{8m\Delta
x_{0}^{4}}\left[\left(x-x_{0,j}\right)^{2}-\left(x'-x_{0,k}
\right)^{2}\right]\label{alfaa}\\
\alpha_{n,\pm}^{j,k}(x,x')\simeq
\alpha_{0}^{j,k}(x,x')\mp\frac{1}{\hbar}ma_{n}t(x-x').\qquad\label{alfab}
\end{eqnarray}
As for $\alpha_{0}^{j,k}(x,x')$, the values of the parameters used
in the paper give $\max\left\{\alpha_{0}^{j,k}(x,x')\right\}\leq\
10^{-3}$, and the phase factors in the second and third terms of
Eq. (\ref{roredxx'}) will only depend on $ma_{n}t(x-x')/\hbar$
which takes into account the optical SG effects in the atomic
momentum distribution.

Taking into account all these approximations, the Eq.
(\ref{roredxx'}) may be written as the product of the initial
spatial density matrix and a decoherence factor,

\begin{equation}\label{rod}
\hat{\rho}(x,x';t)\simeq \hat{\rho}(x,x';0)D(x,x';t)
\end{equation}
where
\begin{eqnarray}
\hat{\rho}(x,x';0)=\frac{1}{2\delta\sqrt{2\pi}\Delta
x_{0}}\qquad\qquad\qquad\qquad\qquad\qquad\nonumber\\
\times\sum_{j,k=1}^{2} \exp\left\{-\frac{1}{4\Delta x_{0}^{2}}
\left[\left(x-x_{0,j}\right)^{2}+\left(x'-x_{0,k}\right)^{2}\right]\right\},\,\,\label{roredxx'a}
\end{eqnarray}
\begin{eqnarray}
D(x,x';t)=\sum_{n=0}^{\infty}\left\{\left(A_{n,n}+B_{n,n}\right)
\cos\left[k\left(x-x'\right)\Omega_{n} t \right]\right.\nonumber\\
\left.-i\left(A_{n,n}-B_{n,n}\right)\sin\left[k\left(x-x'\right)\Omega_{n}
t \right]\right\}+F,\qquad\label{decfac}
\end{eqnarray}
and
\begin{equation}\label{omega}
\Omega_{n}=\Omega\sqrt{\frac{n+1}{<n>+1}},\qquad
\Omega=\varepsilon\sqrt{<n>+1}. \qquad
\end{equation}
Expression (\ref{rod}) is of easy reading and, at the same time, a
very good approximation. In fact, all the figures describing the
decoherence effects presented below, has been obtained from this
equation, but actually they cannot be distinguished from those
related to the exact expression (\ref{roredxx'}).

Expression (\ref{decfac}) is suitable also to explicate the
relation between the field coherence properties and the one-sided
atomic deflection \cite{cus} in the optical SG effect. A mean
value of $\hat{p}$ different from zero requires an imaginary part
of the density operator. For an initial real density operator as
in our case this implies, for $t>0$, $Im[D(x,x';t)]\neq0$, that is
$A_{nn}\neq B_{nn}$. In fact, it is not difficult to show that
\begin{equation}\label{pm}
<\hat{p}>(t)=\hbar k
\sum_{n=0}^{\infty}\left(B_{n,n}-A_{n,n}\right)\Omega_{n}t.
\end{equation}
Looking at Eq.s (\ref{coeffAnn}) and (\ref{coeffBnn}), we conclude
that the atomic center of mass may acquire an increasing momentum
along the cavity axis if its qubit interacts with a field that
owns coherence properties, that is, when $c_{n,n'}$ is not
proportional to $\delta_{n,n'}$.
\section{\label{sec:level2} Some examples of field statistics}
\begin{figure}
\resizebox{0.95\columnwidth}{!}{%
  \includegraphics{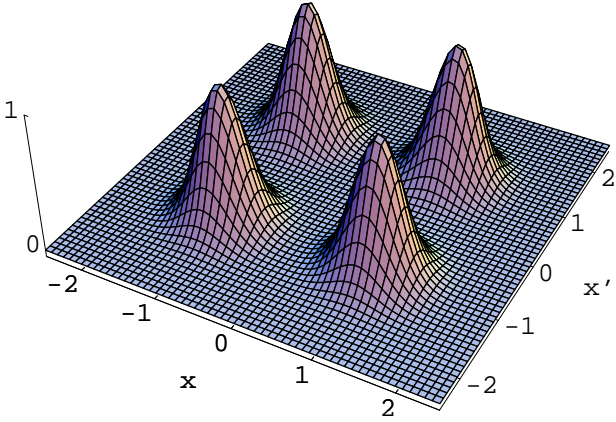}}
\caption{This figure shows the reduced density matrix
$\hat{\rho}(x,x';0)$ of Eq. (\ref{roredxx'a}) and represents the
initial condition for all the cases considered below. It describes
the initial atomic populations (near the $x=x'$ direction) and
coherences (near the $x=-x'$ direction) for the position degree of
freedom along the cavity axis. The values of the parameters are:
$m=10^{-25}$ kg, $\varepsilon=10^5 sec^{-1}$,
$\lambda=0.6\times10^{-2}$ meters, $\Delta x_{0}=\lambda/{100}$,
$x_{0,1}=-\lambda/20$, $x_{0,2}=\lambda/20$. The positions $x,x'$
are considered in units of $|x_{0,1}|$.} \label{fig1}
\end{figure}
In this section the behavior of the atomic spatial coherences will
be studied considering some examples of field states. To this end,
it is sufficient to specialize the coefficients $A_{n,n}, B_{n,n}$
and $F$ of Eq.s (\ref{coeffAnn}-\ref{coeffF}), according to the
specific initial field state, and then use Eq. (\ref{decfac}). The
spatial initial condition for all the cases here considered is
given by $\hat{\rho}(x,x';0)$ of Eq. (\ref{roredxx'a}), and shown
in Fig. \ref{fig1}.  Under appropriate conditions \cite{zu1}, the
two peaks near the $x=x'$ direction  evolve towards the atomic
populations (the two possible atomic positions) for the
translational degree of freedom along the cavity axis, while the
peaks near the orthogonal direction, $x=-x'$, undergo a quenching
when the system evolves towards a classical behavior.\\
\subsection{\label{sec:level2} Incoherent field states}
From our particular point of view, a common interesting feature of
the incoherent field states is the equality, for each $n$, of the
coefficients $A_{n,n}$ and $B_{n,n}$ in Eq. (\ref{decfac}), that
implies the reality of $D(x,x';t)$ and, in our case, of
$\hat{\rho}(x,x';t)$.
\subsubsection{\label{sec:level3}  Thermal state}
\begin{figure}
\resizebox{0.95\columnwidth}{!}{%
  \includegraphics{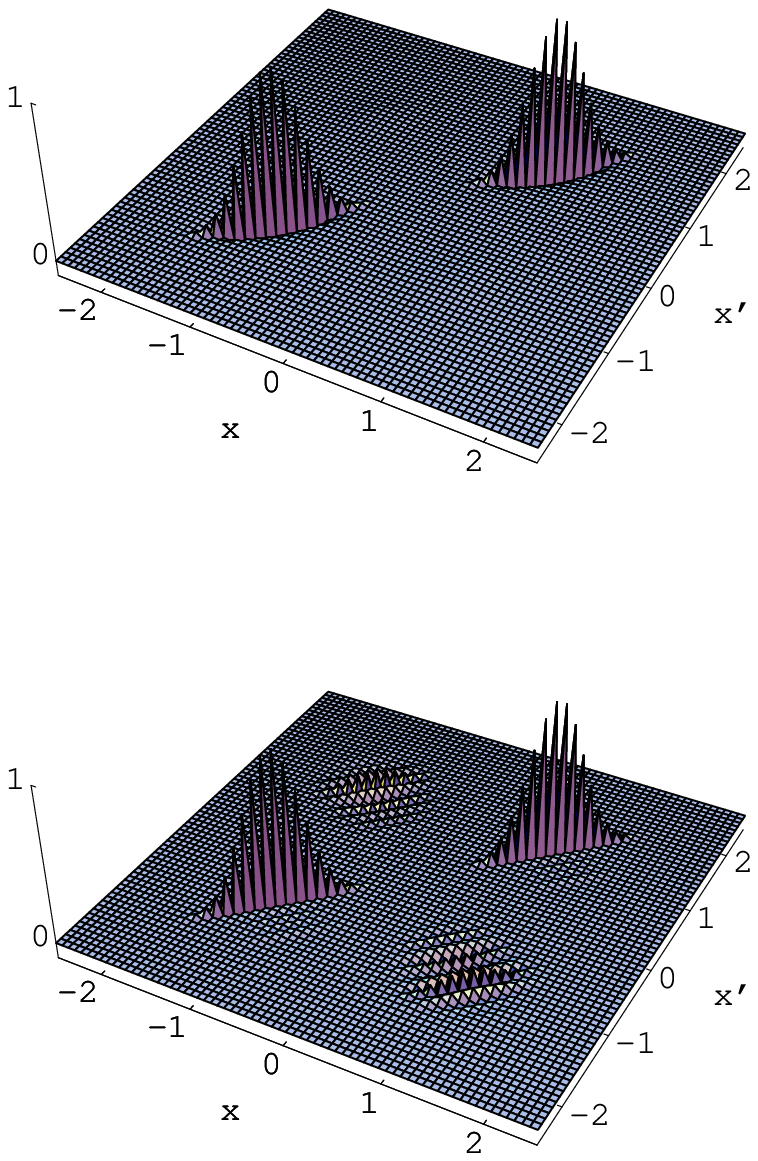}}
 \caption{Position density operator $\hat{\rho}(x,x';t)$ of Eq. (\ref{rod})
 at time $t=100$ $\Omega^{-1}$ (upper) and
 $t=1000$ $\Omega^{-1}$ (lower), when the field is in
 the thermal state (\ref{rotherm}), and the decoherence factor
 $D(x,x';t)$ is given by Eq. (\ref{DT}). The
 parameter values of the cavity and the qubit are:
 $T=200 ^{\circ}K$, corresponding to $<n>$ $ \simeq 82.76$ and $\gamma=\pi/2$.
 The other parameters are as in Fig.1.}
\label{fig2}
\end{figure}

A cavity field in a thermal state at the temperature $T$ is
described by the normalized Boltzmann factor,
\begin{eqnarray}\label{rotherm}
\hat{\rho}_{f}(0)=\frac{e^{-\beta_{B}\hbar\omega\hat{a}^{\dag}\hat{a}}}{Tr(e^{-\beta_{B}\hbar\omega\hat{a}^{\dag}\hat{a}})}
\qquad\qquad\qquad\qquad\nonumber\\
=\left(1-e^{-\beta_{B}\hbar\omega}\right)\sum_{n=0}^\infty
{e^{-n\beta_{B}\hbar\omega}\ket{n}\bra{n}}
\end{eqnarray}

where $\beta_{B}=1/k_{B}T$ and $k_{B}$ is the Boltzmann constant.
Comparison with Eq. (\ref{rof}) gives
\begin{equation}\label{cnnt}
c_{n,n'}=\left(1-e^{-\beta_{B}\hbar\omega}\right)e^{-n\beta_{B}\hbar\omega}\delta_{n,n'}.
\end{equation}
Consequently, for the coefficients $A_{n,n}$, $B_{n,n}$ and $F$ of
Eq.s (\ref{coeffAnn}), (\ref{coeffBnn}) and (\ref{coeffF}) we have
\begin{eqnarray}
A_{n,n}= B_{n,n}=\frac{1}{2}\left(1-e^{-\beta_{B}\hbar\omega}\right)e^{-n\beta_{B}\hbar\omega}\qquad\qquad\nonumber\\
\times\left\{\cos^{2}(\gamma/2)+
e^{-\beta_{B}\hbar\omega}\sin^{2}(\gamma/2)\right\}\qquad\label{ab}\\
F=\sin^{2}(\gamma/2)\left(1-e^{-\beta_{B}\hbar\omega}\right).\qquad\qquad\qquad\qquad\label{f}
\end{eqnarray}
As said, the equality $A_{n,n}= B_{n,n}$ implies that the
decoherence factor,
\begin{eqnarray}\label{DT}
D(x,x';t)= \sin^{2}(\gamma/2)\left(1-e^{-\beta_{B}\hbar\omega}\right)\qquad\qquad\qquad\nonumber\\
+\left(1-e^{-\beta_{B}\hbar\omega}\right)\left[\cos^{2}(\gamma/2)+
e^{-\beta_{B}\hbar\omega}\sin^{2}(\gamma/2)\right]\nonumber\\
\times\sum_{n=0}^{\infty}\cos\left[k\left(x-x'\right)\Omega_{n}t\right]e^{-n\beta_{B}\hbar\omega}.\qquad\qquad\qquad
\end{eqnarray}
is a real function. The corresponding density matrix
$\hat{\rho}(x,x';t)$ $=\hat{\rho}(x,x';0)$ $D(x,x';t)$ is shown in
Fig. \ref{fig2}. The entanglement with the field-qubit environment
causes a spatial coherence decay (upper), followed by a very
partial revival (lower), because of the discreteness of the states
to which the environment can accede. It is evident that by
increasing $T$ the revivals should be postponed in time.

One may also analyze the coherence behavior by looking at the
function $\hat{\rho}(x,x';t)$ along the $x'=-x$ direction. Fig.
\ref{fig3} shows a temporal sequence of the function
$\hat{\rho}(x'=-x;t)$ for the same thermal case. In what follows,
the spatial coherences will be only analyzed in terms of function
of this kind.
\subsubsection{\label{sec:level2}  Coherent state with random
phase}
A mixture of coherent states with random phases is given by
\begin{eqnarray}\label{rofcr}
\hat{\rho}_{f}(0)=\frac{1}{2\pi}\int_{0}^{2\pi}d\vartheta\ket{\alpha}\bra{\alpha}\qquad\qquad\qquad\nonumber\\
=e^{-|\alpha|^{2}}\sum_{n=0}^\infty
\frac{\alpha^{2n}}{n!}\ket{n}\bra{n}.\qquad\qquad
\end{eqnarray}
In this case we have
\begin{equation}\label{cnntcr}
c_{n,n'}=\frac{\alpha^{2n}}{n!}e^{-|\alpha|^{2}}\delta_{n,n'}
\end{equation}
and using Eq.s (\ref{coeffAnn}-\ref{coeffF}) we obtain
\begin{eqnarray}
A_{n,n}=B_{n,n}=\frac{|\alpha|^{2n}}{2n!}e^{-|\alpha|^{2}}\left[\cos^{2}\frac{\gamma}{2}
+\frac{|\alpha|^{2}}{n+1}\sin^{2}\frac{\gamma}{2}\right]\label{abcr}\\
F=e^{-|\alpha|^{2}}\sin^{2}(\gamma/2).\qquad\qquad\qquad\qquad\label{fc}
\end{eqnarray}
The decoherence function,
\begin{eqnarray}\label{DCR}
D(x,x';t)=e^{-|\alpha|^{2}}\sin^{2}(\gamma/2)
+e^{-|\alpha|^{2}}\sum_{n=0}^{\infty}\frac{|\alpha|^{2n}}{n!}\nonumber\\
\times\left(\cos^{2}\frac{\gamma}{2}
+\frac{|\alpha|^{2}}{n+1}\sin^{2}\frac{\gamma}{2}\right)
\cos\left[k\left(x-x'\right)\Omega_{n}t\right]
\end{eqnarray}
is equal to real part of the same function relative to the
coherent state (see sec. 4.2.1). Consequently, Fig. \ref{fig5}
describes also this case.
\subsubsection{\label{sec:level2}  Fock state}
Finally, for a Fock state
\begin{equation}\label{rofn}
\hat{\rho}_{f}(0)=\ket{n_{0}}\bra{n_{0}},
\end{equation}
it is $c_{n,n'}=\delta_{n,n_{0}}\delta_{n',n_{0}}$. Consequently,
one has
\begin{equation}\label{abfo}
A_{n,n}=B_{n,n}=\frac{1}{2}\left[\delta_{n,n_{0}}\cos^{2}\frac{\gamma}{2}+
\delta_{n,n_{0}-1}\sin^{2}\frac{\gamma}{2}\right],
\end{equation}
\begin{equation}\label{ffo}
F=\delta_{n_{0},0}\sin^{2}(\gamma/2)\qquad\qquad\qquad\qquad
\end{equation}
and
\begin{eqnarray}\label{DN}
D(x,x';t)=\cos^{2}(\gamma/2)
\cos\left[k\left(x-x'\right)\Omega_{n_{0}}t\right]\qquad\nonumber\\
+\sin^{2}(\gamma/2)\cos\left[k\left(x-x'\right)\Omega_{n_{0}-1}t\right]
+\sin^{2}(\gamma/2)\delta_{n_{0},0}.
\end{eqnarray}
\begin{figure}
\resizebox{0.95\columnwidth}{!}{%
  \includegraphics{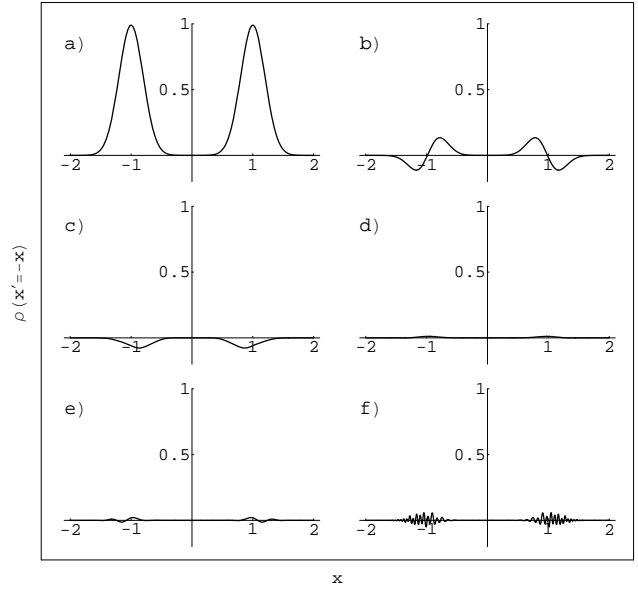}}
 \caption{This figure describes in a simple way the decoherence effect (followed by a partial revival,
 curve (f)), for the thermal state (\ref{rotherm}). It represents the function
 $\hat{\rho}(x,x';t)$ defined by Eq.s (\ref{rod}), (\ref{roredxx'a}) and (\ref{DT}), along the
 $x'=-x$ direction. From (a) to (f) time is
$0,3,10,100,200,500$, in units of $\Omega^{-1}$. The other
parameters are as in Fig. 2} \label{fig3}
\end{figure}
The behavior of $\hat{\rho}(x'=-x;t)$ for the number case, shown
in Fig. \ref{fig4}, outlines how the coherences survive at all
times if the exchange of energy (and information) between the
system of interest and the environment follows a few paths.
\subsection{\label{sec:level2}  Field states with phase properties}
When the field density matrix
$c_{n,n'}=\bra{n}\hat{\rho}_{f}(0)\ket{n'}$ of Eq. (\ref{rof}) is
not diagonal, the coefficients $A_{n,n}$ and $B_{n,n}$ of Eq.
(\ref{decfac}) may be different between them. $D(x,x';t)$ and
$\hat{\rho}(x,x';t)$ are in general complex functions.
\subsubsection{\label{sec:level2}  Coherent state}
For an initial coherent state of the cavity field,
\begin{equation}\label{rofc}
\hat{\rho}_{f}(0)=e^{-|\alpha|^{2}}\sum_{n,n'=0}^\infty
\frac{\alpha^{n}(\alpha^{\ast})^{n'}}{\sqrt{n!n'!}}\ket{n}\bra{n'},
\end{equation}
we have
\begin{equation}\label{cnntc}
c_{n,n'}=\frac{\alpha^{n}(\alpha^{\ast})^{n'}}{\sqrt{n!n'!}}e^{-|\alpha|^{2}};\qquad
\alpha=|\alpha|e^{i\theta}
\end{equation}
and
\begin{eqnarray}
A_{n,n}=\frac{|\alpha|^{2n}}{2n!}e^{-|\alpha|^{2}}\left|\cos\frac{\gamma}{2}
+\frac{|\alpha|
e^{i(\theta+\phi)}}{\sqrt{n+1}}\sin\frac{\gamma}{2}\right|^{2}\label{ac}\\
B_{n,n}=\frac{|\alpha|^{2n}}{2n!}e^{-|\alpha|^{2}}\left|\cos\frac{\gamma}{2}
-\frac{|\alpha|
e^{i(\theta+\phi)}}{\sqrt{n+1}}\sin\frac{\gamma}{2}\right|^{2}\label{bc},
\end{eqnarray}
while coefficient $F$ is still given by Eq. (\ref{fc}). The
decoherence factor is
\begin{eqnarray}\label{DC}
D(x,x';t)=e^{-|\alpha|^{2}}\sin^{2}(\gamma/2)
+e^{-|\alpha|^{2}}\sum_{n=0}^{\infty}\frac{|\alpha|^{2n}}{n!}\nonumber\\
\times\left(\cos^{2}\frac{\gamma}{2}
+\frac{|\alpha|^{2}}{n+1}\sin^{2}\frac{\gamma}{2}\right)
\cos\left[k\left(x-x'\right)\Omega_{n}t\right]\nonumber\\
-2i|\alpha|e^{-|\alpha|^{2}}\sin \gamma \cos(\theta+\phi)\qquad\qquad\qquad\nonumber\\
\times\sum_{n=0}^{\infty}\frac{|\alpha|^{2n}}{n!\sqrt{n+1}}
\sin\left[k\left(x-x'\right)\Omega_{n}t\right].\qquad\qquad
\end{eqnarray}
whose real part coincides with the decoherence function
(\ref{DCR}). We recall that, for $<n>$ sufficiently large (as it
is in our case) and $\gamma=\pi/2$, the phase relation between the
qubit and the field may lead to two opposite behaviors. For
$\theta+\phi=\pi/2$ one has $A_{n,n}\simeq B_{n,n}$, $D(x,x';t)$
and $\hat{\rho}(x,x';t)$ are real, and $<\hat{p}>(t)=0$. On the
contrary, for $\theta+\phi=0$, or $\pi$ the decoherence factor is
complex and a one-sided deflection characterizes the optical SG
effect. In fact one has $B_{n,n}=0$, or $A_{n,n}=0$, respectively
and the atom is scattered to left or to the right (see Eq.
(\ref{pm})). This is the so called trapping condition, after the
fact that the qubit population is approximately trapped to the
initial value, with the consequent quenching of the Rabi
oscillations \cite{zz,cus}. (For the relation between the
interference phenomenon of the Rabi oscillations and the selective
deflection in the optical SG model see Ref. \cite{tum}). For the
case $\theta+\phi=0$ and $\gamma=\pi/2$, the behavior of the real
and imaginary part of $\hat{\rho}(x'=-x;t)$ is shown in Fig.s
\ref{fig5} and \ref{fig6}, respectively.
\subsubsection{\label{sec:level2}  Phase state}
The exact coherent trapping condition \cite{Cir,cus} may be
achieved when the cavity field is in an eigenstate of the
Susskind-Glogower phase operator \cite{SussCarr,Cir}
$\exp{(i\hat{\theta})}=(\hat{n}+1)^{-1/2}\hat{a}$. In this case
the field is
\begin{equation}\label{roff}
\hat{\rho}_{f}(0)=(1-|z|^{2})\sum_{n,n'=0}^\infty
z^{n}(z^{\ast})^{n'}\ket{n}\bra{n'}.
\end{equation}
and
\begin{equation}\label{cnntf}
c_{n,n'}=(1-|z|^{2})z^{n}(z^{\ast})^{n'}.
\end{equation}
The coherent trapping takes on for particular relations between
the field and the qubit parameters. In fact, it is required that
\begin{figure}
\resizebox{0.95\columnwidth}{!}{%
  \includegraphics{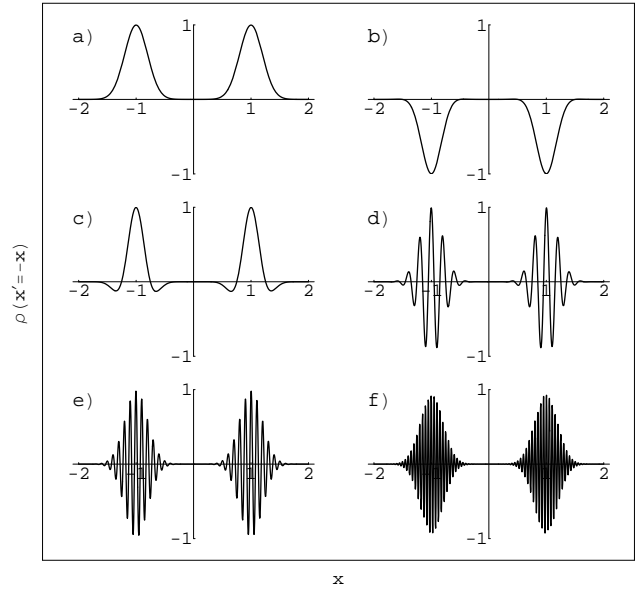}}
 \caption{Same quantity of Fig. 3 when the cavity field is
 in the Fock state (\ref{rofn}), with $n_{0}=83$, and $\gamma=\pi/2$.
From (a) to (f) time is $0,5,10,50,100,200$, in units of
$\Omega^{-1}$. The other parameters are as in Fig. 1.}
\label{fig4}
\end{figure}
\begin{figure}
\resizebox{0.95\columnwidth}{!}{%
  \includegraphics{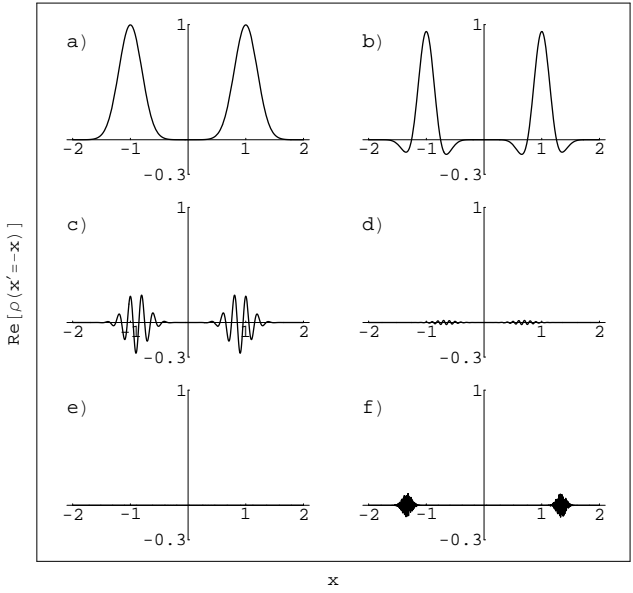}}
 \caption{Real part of $\hat{\rho}(x'=-x;t)$ when the cavity field is
 in the coherent state (\ref{rofc}) with $|\alpha|^{2} = <n> =
 82.76$, and the field-qubit
 system is in the trapping configuration, $\theta+\phi=0$ and
 $\gamma=\pi/2$. From (a) to (f) time is
$0,10,50,100,500,1200$, in units of $\Omega^{-1}$. The other
parameters are as in Fig. 1. This figure describes also the
behavior of $\hat{\rho}(x'=-x;t)$ for the coherent state with
random phase, see Eq.s (\ref{DCR}) and (\ref{DC}).} \label{fig5}
\end{figure}
\begin{equation}\label{zgamma}
z=e^{i\theta}\cot(\gamma/2),\qquad\qquad\pi/2<\gamma\leq\pi,
\end{equation}
and $\theta+\phi=0$ or $\pi$.

By using Eq.s (\ref{coeffAnn}-\ref{coeffF}) and (\ref{cnntf}) we
obtain
\begin{eqnarray}
A_{n,n}= \frac{1}{2}(1-|z|^{2})|z|^{2n}\{\cos^{2}(\gamma/2)\qquad\qquad\qquad\qquad\nonumber\\
+2|z|\cos(\theta+\phi)\cos(\gamma/2)\sin(\gamma/2)+|z|^{2}\sin^{2}(\gamma/2)\}\qquad\label{af}\\
B_{n,n}=\frac{1}{2}(1-|z|^{2})|z|^{2n}\{\cos^{2}(\gamma/2)\qquad\qquad\qquad\qquad\nonumber\\
-2|z|\cos(\theta+\phi)\cos(\gamma/2)\sin(\gamma/2)+|z|^{2}\sin^{2}(\gamma/2)\}\qquad\label{bf}\\
F=\sin^{2}(\gamma/2)(1-|z|^{2})\qquad\qquad\qquad\qquad\label{ff}
\end{eqnarray}
\begin{figure}
\resizebox{0.95\columnwidth}{!}{%
  \includegraphics{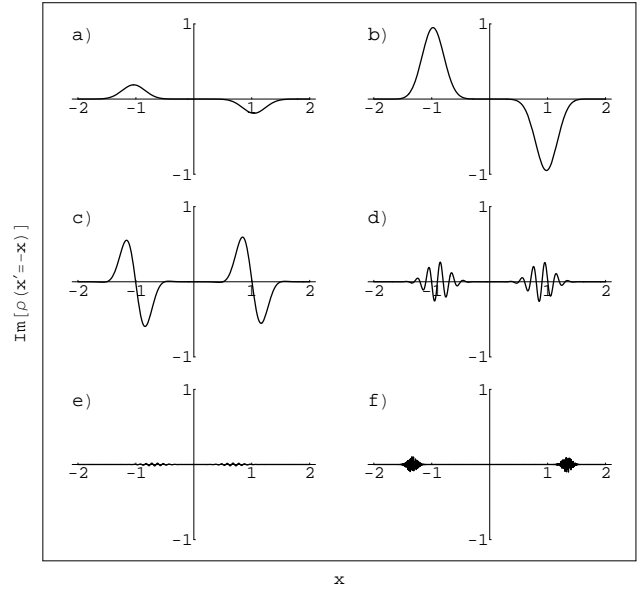}}
 \caption{Imaginary part of $\hat{\rho}(x'=-x;t)$ for the same
 configuration of Fig. 5.
  This figure shows the growing (from zero) of the imaginary part of the atomic position
  coherence, followed by a loss and a partial revival.
 From (a) to (f) time, in
units of $\Omega^{-1}$, is $0.3,3,10,50,100,1200$. The other
parameters are as in Fig. 5} \label{fig6}
\end{figure}
As one can easily see, for $\theta+\phi=0$ ($\theta+\phi=\pi$) one
has $B_{n,n}=0$ ($A_{n,n}=0$). The previous consideration on the
interplay between coherent trapping, quenching of the Rabi
oscillations and selective atomic deflection in this case hold
exactly.
\begin{figure}
\resizebox{0.95\columnwidth}{!}{%
  \includegraphics{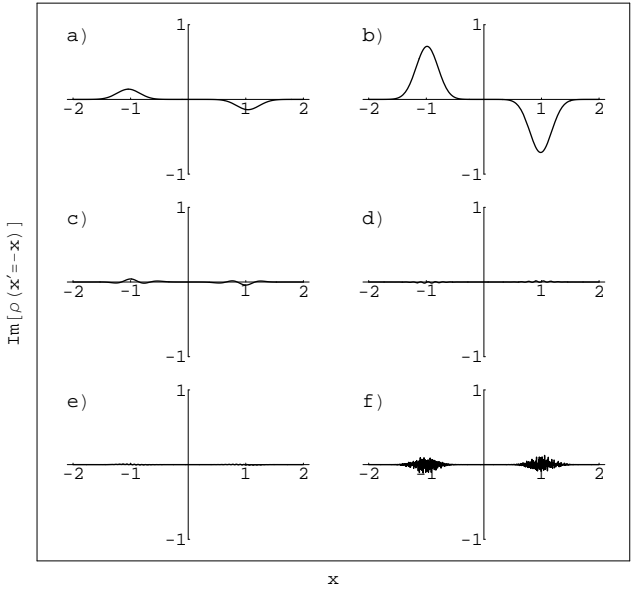}}
 \caption{Imaginary part of $\hat{\rho}(x'=-x;t)$ for the
 phase state (\ref{roff}) in the trapping configuration ($\theta+\phi=0$),
 with $\gamma=0.5019115\pi$, for which it is $<n>=82.76$. The other
 parameters are as in Fig. 6}
\label{fig7}
\end{figure}
It is also to note that the photon distribution of state
(\ref{roff}) follows a geometric law,
\begin{equation}\label{pn}
P_{n}= \bra{n}\hat{\rho}_{f}(0)\ket{n}=(1-|z|^{2})|z|^{2n}
=\frac{<n>^{n}}{(1+<n>)^{n+1}}
\end{equation}
with
\begin{equation}\label{nm}
<n>=
\frac{|z|^{2}}{(1-|z|^{2})}=\frac{\cot^{2}{(\gamma/2)}}{(1-\cot^{2}{(\gamma/2}))}.
\end{equation}
The pure state (\ref{roff}) owns the same photon statistics of the
mixed thermal state (\ref{rotherm}). For this reason, the behavior
of $Re[\hat{\rho}(x'=-x;t)]$ is very similar to the thermal case
and, for the phase state, we report in Fig. \ref{fig7}
only $Im[\hat{\rho}(x'=-x;t)]$. \\
\begin{figure}
\resizebox{0.95\columnwidth}{!}{%
  \includegraphics{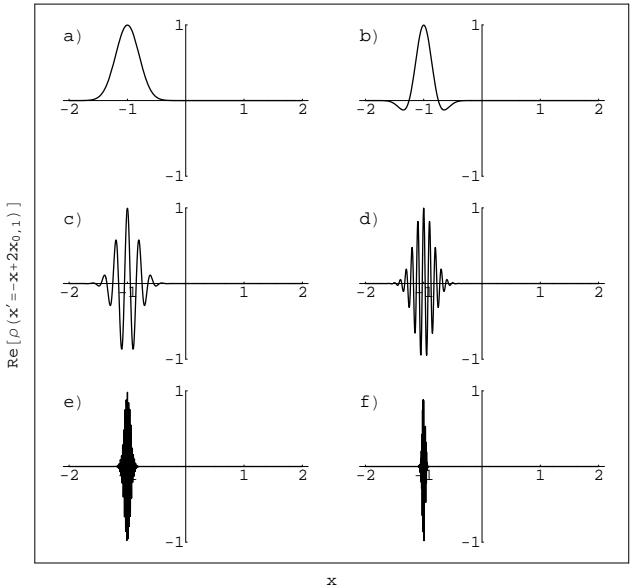}}
\caption{Real part of $\hat{\rho}(x'=-x+2x_{0,1};t)$ for the
coherent state with $|\alpha|^{2} = 82.76$, in the trapping
configuration, $\theta+\phi=0$) and $\gamma=\pi/2$. We recall that
the position variable $x$ is considered in units of $|x_{0,1}|$.
From (a) to (f) time is $0,10,50,100,500,1000$, in units of
$\Omega^{-1}$. The other parameters are as in Fig. 1} \label{fig8}
\end{figure}
\begin{figure}
\resizebox{0.95\columnwidth}{!}{%
  \includegraphics{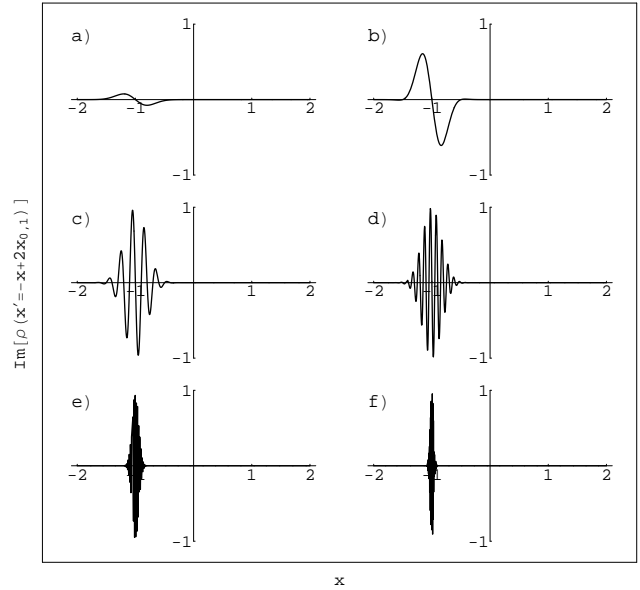}}
\caption{Imaginary part of $\hat{\rho}(x'=-x+2x_{0,1};t)$ for the
coherent state of Fig. 6. From (a) to (f) time is
$1,10,50,100,500,1000$, in units of $\Omega^{-1}$. The other
parameters are as in Fig. 8} \label{fig9}
\end{figure}
Comparing the standard deviations of the geometric distribution,
$\Delta n\simeq <n>$ for large $n$, and of the Poissonian law,
$\Delta n = \sqrt{<n>}$, we may conclude that for the same value
of $<n>$ the interaction with a thermal (or phase) state takes
place following more alternatives with respect to the coherent
case. As a consequence, coherences fall off more quickly in the
first case, as one can see comparing Fig.s \ref{fig3} and
\ref{fig5} for the real part of the density matrix, and Fig.s
\ref{fig7} and \ref{fig6} for the imaginary part.

It is to notice that the coherence revivals of Fig.s \ref{fig5},
\ref{fig6} and \ref{fig7} are displayed for times which are
borderline with respect to the validity of the model.
\section{\label{sec:level2} Decoherence effects on local coherences}
Let us come back  briefly to the link between the imaginary part
of the spatial density matrix and the atomic momentum. It may be
interesting to note that this link involves some constraints to
the decoherence process of the imaginary part of
$\hat{\rho}(x,x';t)$, at least in some regions of of the plane
$(x,x')$. Consider again Fig.\ref{fig1}. The two peaks along the
$x'=-x$ direction describe spatial coherences at the initial time
(which we may denote as nonlocal). As we have seen, the
entanglement with the field-qubit environment causes a decoherence
of $\hat{\rho}(x,x';t)$ in terms of a time decay of these two
peaks. Because of the discreteness of the environment states
density, the considerations brought in the previous section only
indicate a propensity to a classical landing, as the figures of
the previous section show.

As for the other two peaks located in the orthogonal direction,
one could say that they describe the atomic populations, that is,
the two possible atomic initial position \cite{zu1}. Actually,
when $x'$ is near to but not coincident with $x$, the function
$\hat{\rho}(x,x';0)$ describes coherences (which we may denote as
local). An intriguing question concerns the time evolution of
these two peaks. First of all, the asymptotic form of the real
part should give the genuine populations, however, for continuity
reasons, the local coherences cannot abruptly vanish. The Fig.
\ref{fig8} reports $Re[\hat{\rho}(x,x';t)]$ along a direction
parallel to $(x'=-x)$, thought one of the two maxima located in
$x_{0,i}$. One may note how $x_{0,1}$ asymptotically behaves as an
accumulation point for the local coherences.

As far as the imaginary part is concerned, also in this case a
decay similar to what happens for the nonlocal coherences cannot
be granted. In fact, as seen in sec. 3, the mean value of the
atomic momentum is connected to the $Im[\hat{\rho}(x,x';t)]$. More
precisely, since in our case $Im[\hat{\rho}(x,x';0)]=0$ (and
$<\hat{p}>(0)=0$), an increasing mean value of $<\hat{p}>$
requires a non vanishing imaginary decoherence factor, as one may
see from Eq.s (\ref{pm}) and (\ref{decfac}). We find, in fact,
that the local imaginary coherences are not subjected to an
asymptotical decay, as Fig. \ref{fig9} shows. Since
$<\hat{p}>(0)=0$, the function $Im[\hat{\rho}(x'=-x+2x_{0,1};t)]$
grows up from zero, and for $t$ sufficiently large, its amplitude
behaves similarly to the real case of Fig. \ref{fig8}, with an
essential difference: at the accumulation point $x_{0,1}$,
$Im[\hat{\rho}(x'=-x+2x_{0,1};t)]$ is zero, with an increasing
slope which accounts for the atomic acceleration.
\section{\label{sec:level2} Conclusions}
The optical Stern-Gerlach model is a useful tool for analyze
disentanglement and decoherence processes that involve
environments with a few degrees of freedom both for discrete and
continuous variables. It consists essentially of a Jaynes-Cummings
two level atom which entangles with its translational dynamics
along the cavity axis. Under the environmental action of the
Jaynes-Cummings qubit we analyze the decoherence process of the
atomic translation, considered as system of interest. We find that
the decoherence features may be encoded into a decoherence
function $D(x,x';t)$. In fact, for the more general one mode state
(pure or mixed) of the cavity field, the atomic spatial density
matrix may be factorized as $\rho(x,x';t)=\rho(x,x';0)$
$D(x,x';t)$, where the decoherence factor $D(x,x';t)$ depends on
the statistics and on the phase properties of the field. The field
statistics affects $D(x,x';t)$ through the density of quantum
paths followed in the interchange of energy and information
between the subsystems, while the phase properties of the field
(more precisely, the qubit-field phase relations) are related to
the imaginary part of $D(x,x';t)$ and to the atomic kinematics.

We find useful to distinguish the local coherences
$\hat{\rho}(x'=-x+2x_{0,i};t)$ in the neighborhoods of the two
population peaks, $x\simeq x_{0,i}\simeq x'$ ($i=1,2$), from the
non local ones $\hat{\rho}(x'=-x;t)$, along the direction $x'=-x$
(see Fig. 1). As expected, $\hat{\rho}(x'=-x;t)$ asymptotically
vanishes, with decay times depending on the environment states
density, and consequently, on the field statistics (because of the
discreteness of these states, the non local coherences are
subjected to a partial revival at larger times). On the contrary,
the local coherences cannot abruptly vanish, since their real part
has to ensure the asymptotic survival of the populations.
Surprisingly, also the imaginary local coherences do survive for
increasing values of the atomic momentum, and the neighborhoods of
the two population peaks asymptotically became a sort of
accumulation points for the local coherences. In our opinion,
these peculiarities of the local coherences deserve further
analysis.

%

%

%
%

% ----------------------------------------------------------------
%

%
%
\end{document}